\documentclass[a4paper,11pt]{article}
\pdfoutput=1
\usepackage{jheppub} 
\usepackage[english]{babel}
\usepackage{multirow}
\usepackage{amsmath,amssymb}
\usepackage{graphicx}
\usepackage{caption}
\usepackage{subcaption}
\usepackage{color}
\usepackage{url}
\usepackage{slashed}
\usepackage{footmisc}
\usepackage{hyperref}
\usepackage{ragged2e}
\DeclareCaptionJustification{justified}{\justifying}
\newcommand{\blue}[1]{\textcolor{blue}{#1}}

\usepackage{graphicx}

\begin{document}

\title{Di-Higgs production in the $4b$ channel and
Gravitational Wave complementarity}

\author[a]{Alexandre Alves}
\author[b,c]{Dorival Gon\c{c}alves} 
\author[d]{Tathagata Ghosh}
\author[e]{Huai-Ke Guo}
\author[e]{Kuver Sinha}
\affiliation[a]{Departamento de F\'isica, Universidade Federal de S\~ao Paulo, UNIFESP, Diadema, Brazil}
\affiliation[b]{PITT PACC, Department of Physics and Astronomy, University of Pittsburgh, 3941 O'Hara St., Pittsburgh, PA 15260, USA}
\affiliation[c]{Department of Physics, Oklahoma State University, Stillwater, OK, 74078, USA}
\affiliation[d]{Department of Physics \& Astronomy, University of Hawaii, Honolulu, HI 96822, USA}
\affiliation[e]{Department of Physics and Astronomy, University of Oklahoma, Norman, OK 73019, USA}
\emailAdd{aalves@unifesp.br}
\emailAdd{dorival@okstate.edu}
\emailAdd{tghosh@hawaii.edu}
\emailAdd{ghk@ou.edu}
\emailAdd{kuver.sinha@ou.edu}

\preprint{PITT-PACC-1904,~OSU-HEP-19-06}

\abstract{
We present a complementarity study of gravitational waves and double Higgs production in the $4b$ channel, exploring the gauge singlet scalar extension of the SM. This new physics extension serves as a simplified benchmark model that realizes a strongly first-order electroweak phase transition necessary to generate the observed baryon asymmetry in the universe. In calculating the signal-to-noise ratio of the gravitational waves, we incorporate
the effect of the recently discovered significant suppression of the gravitational wave signals from sound waves for strong phase 
transitions, make sure that supercooled phase transitions do complete and adopt a bubble wall velocity that is consistent with a successful electroweak baryogenesis by solving the velocity profiles of the plasma. The high-luminosity LHC sensitivity to the singlet scalar extension of the SM is estimated using a shape-based analysis of the invariant $4b$ mass distribution. We find that while the region of parameter space giving detectable gravitational waves is shrunk due to the new gravitational wave simulations, the qualitative complementary role of gravitational waves and collider searches remain unchanged. 
}

\maketitle

\section{Introduction}
The first direct detection of the gravitational waves (GW) by the LIGO and Virgo collaborations~\cite{Abbott:2016blz} has triggered a revived interest in using the stochastic GW from the electroweak phase transition (EWPT) to learn more about particle physics, in particular, to probe possible hints of physics beyond the standard model (BSM). A stochastic background of GW can be produced from a cosmological first-order phase transition. The EWPT is required to be  first-order to provide a non-equilibrium environment for generating the observed baryon asymmetry in the universe, in the framework of electroweak baryogenesis (EWBG) (see~\cite{Morrissey:2012db} for a recent review). EWBG is one of the popular
mechanisms for solving the long-standing baryon asymmetry problem. In this framework, BSM physics provides new sources of CP-violation~\cite{Buckley:2015vsa,Kobakhidze:2015xlz,Goncalves:2018agy} and facilitates strong first-order EWPT, both of which cannot be achieved within the SM, though it is possible to violate the baryon number in the SM through the weak Sphaleron process. Therefore, GW measurements can provide a new window to BSM physics.

A relevant theoretical benchmark construction is the so-called ``xSM'', which is a minimal extension of the SM by adding a new gauge singlet scalar. The xSM has been under extensive phenomenological studies due to its simplicity and also been used to study the GW signals due to its ease to accommodate a strong first-order EWPT~\cite{Profumo:2007wc,Profumo:2014opa,Kozaczuk:2015owa,Huang:2017jws,Gould:2019qek,Chen:2017qcz,Carena:2018vpt,Jaeckel:2016jlh}. In a previous paper, we have performed a full analysis of the xSM parameter space and identified the 
features of the parameter space that can give detectable gravitational waves~\cite{Alves:2018jsw}. One of the most important results in the GW literature since our previous study is a very recent numerical simulation of GW production, which found a significant deficit of the produced GW signals from sound waves~\cite{Cutting:2019zws}. The suppression arises presumably from the slowing down of expanding broken phase bubbles due to formation of reheated droplets of unbroken phase. The GW production can be reduced by a factor as small as $0.001$ and it has profound implications as all previous studies might have overestimated the GW signal strengths. Hence, in this paper we checked the impact of the findings of Ref.~\cite{Cutting:2019zws} on a part of the xSM parameter space and explore how it affects the complementarity of the future space-based GW experiments and the collider searches at the LHC. As a consequence of the above suppression of GW production, a detectable GW signal generation will require highly supercooled EWPT. Such supercooled EWPT may lead to vacuum energy dominated universe resulting in the universe stuck in a false vacuum. We address this issue with detailed analysis for our GW benchmarks in this paper, in contrast to our previous studies on GW~\cite{Alves:2018jsw,Alves:2018oct}. We also present a hydrodynamics analysis of the fluid velocity profiles to determine bubble wall velocities that are consistent with EWBG, similar to our previous papers. This is important for the EWBG to generate the observed baryon asymmetry successfully. 

In light of above changes, in this work we perform a dedicated study of the resolved $4b$ decay channel of the di-Higgs production to explore the parameter space giving large detectable GW. In previous papers, we have studied the GW complementarity in the $h_2 \to h_1h_1\to \bar{b}b\gamma \gamma$~\cite{Alves:2018oct} and $h_2 \to VV \, (V=W,Z) $~\cite{Alves:2018jsw} channels. In contrast, in this work, we analyze another potentially major channel that can discover $h_2$ at the LHC complementing the collider studies of our previous two papers mentioned above. Our estimate of the LHC sensitivity to this channel benefits from a reliable background estimation from the recent ATLAS search for double Higgs production and decay to $4b$-tagged jets~\cite{Aaboud:2018knk}. There has been a similar $4b$ analysis in the literature in connection with EWPT~\cite{Li:2019tfd} using several benchmarks, which, however, does not include GW analysis.
We find that the High Luminosity LHC (HL-LHC) measurements in the $4b$ channel will be able to probe the xSM parameters space predicting not too small $h_2\to h_1h_1$ branching ratio while LISA can typically detect strong GW even with tiny $h_2\to h_1h_1$ branching ratio thus establishing the complementary roles of colliders and satellite experiments in constraining models predicting first order EWPT.

Before proceeding, we emphasize that the primary goal of this paper is not to investigate the xSM model but to use it as a template to study new physics. The xSM framework by construction is designed to maximally elucidate the physics of phase transitions and double Higgs production in the most elementary setting, exposing the critical issues without being distracted by complications in the Higgs potential. For example, one could perform the same calculations in extended Higgs sectors of greater complexity. The novel features of our work are: (i) a careful analysis of resonant di-Higgs production in the 4$b$ channel; (ii) a thorough examination of the impact of the suppression factor from very recent gravitational wave simulations; and (iii) investigating the possibility of a supercooled EWPT and its implications.

The paper is organized as follows. We first give a brief introduction to the xSM in the Sec.~\ref{sec:model}, followed by a description of the GW calculations in Sec.~\ref{sec:gw}. We present the dedicated $4b$ collider analysis in Sec.~\ref{sec:4b} and conclude in Sec.~\ref{sec:summary}.

\section{\label{sec:model}The Model}

The xSM model is defined by adding a gauge singlet real scalar to the SM with the following scalar sector potential~\cite{Profumo:2007wc,Profumo:2014opa,Huang:2017jws}:
\begin{eqnarray}
  V(H,S) &=& -\mu^2 H^{\dagger} H + \lambda (H^{\dagger}H)^2 
  + \frac{a_1}{2} H^{\dagger} H S  
    + \frac{a_2}{2} H^{\dagger} H S^2 + \frac{b_2}{2} S^2 + \frac{b_3}{3} S^{3} + \frac{b_4}{4}S^4 . \quad 
  \label{eq:v}
\end{eqnarray} 
Here $H^{\text{T}} = (G^+, (v_{\text{EW}} + h + i G^0)/\sqrt{2})$ is the SM Higgs doublet 
and $S=v_s + s$ is the additional singlet scalar. The parameters in this potential are all real. Of these parameters,
two ($\mu$, $b_2$) can be replaced by $v_s$ and $v_{\text{EW}}$ through the minimization conditions
of the scalar fields; another three parameters ($\lambda$, $a_1$, $a_2$) can be replaced by the masses and
mixing angle of the physical scalars $(m_{h_1}, m_{h_2}, \theta)$. The physical scalars are defined by
\begin{eqnarray}
h_1 = c_{\theta} h + s_{\theta} s, \quad \quad
h_2 =-s_{\theta} h + c_{\theta} s,  
  \label{eq:mixing}
\end{eqnarray}
where $h_1$ is identified as the SM Higgs while $h_2$ is a heavier scalar. With this setup, the potential
is fully specified by the following five unknown parameters:
\begin{eqnarray}
\centering
v_s, \quad \quad m_{h_2}, \quad \quad \theta, \quad \quad b_3, \quad \quad b_4 .
\end{eqnarray}
The parameter space defined by the five parameters above can be subjected to broadly two categories of constraints. The first set of constraints comes directly from various theoretical requirements imposed on the scalar potential, including boundedness of the potential from below, the stability of the EW vacuum, and perturbative unitarity of $2 \rightarrow 2$ scattering processes. All the other constraints are phenomenological. Higgs signal strength measurement~\cite{Khachatryan:2016vau} constrains the mixing angle $\theta$: 
$|\sin \theta| < 0.33$ at $95\%$ CL~\cite{Khachatryan:2016vau}. Another set of constraints comes from EW precision measurements such as the oblique $S,T,U$ parameters~\cite{Peskin:1991sw,Hagiwara:1994pw} and correction to the the $W$ boson mass $m_W$~\cite{Lopez-Val:2014jva}. Both the EW precision measurements mentioned above constrain only $(m_{h_2}, \theta)$
at one-loop level, with the $m_W$ measurement providing a more stringent bound~\cite{Lopez-Val:2014jva,Robens:2015gla}. For the details of this model and impact of the constraints used on the parameter space, we refer the reader to our previous paper~\cite{Alves:2018jsw}.
We further note that a successful EWBG also needs additional CP-violation, to fulfill one other
Sakharov condition. However, it is typically very constrained by the stringent EDM limits so that it tends to have a minor effect on the EWPT. For larger CP-violation, which is less constrained by the EDM constraints and negligibly affect the EWPT, see, e.g., Ref.~\cite{Guo:2016ixx}.

\begin{figure}[ht]
  \centering
  \includegraphics[width=0.44\textwidth]{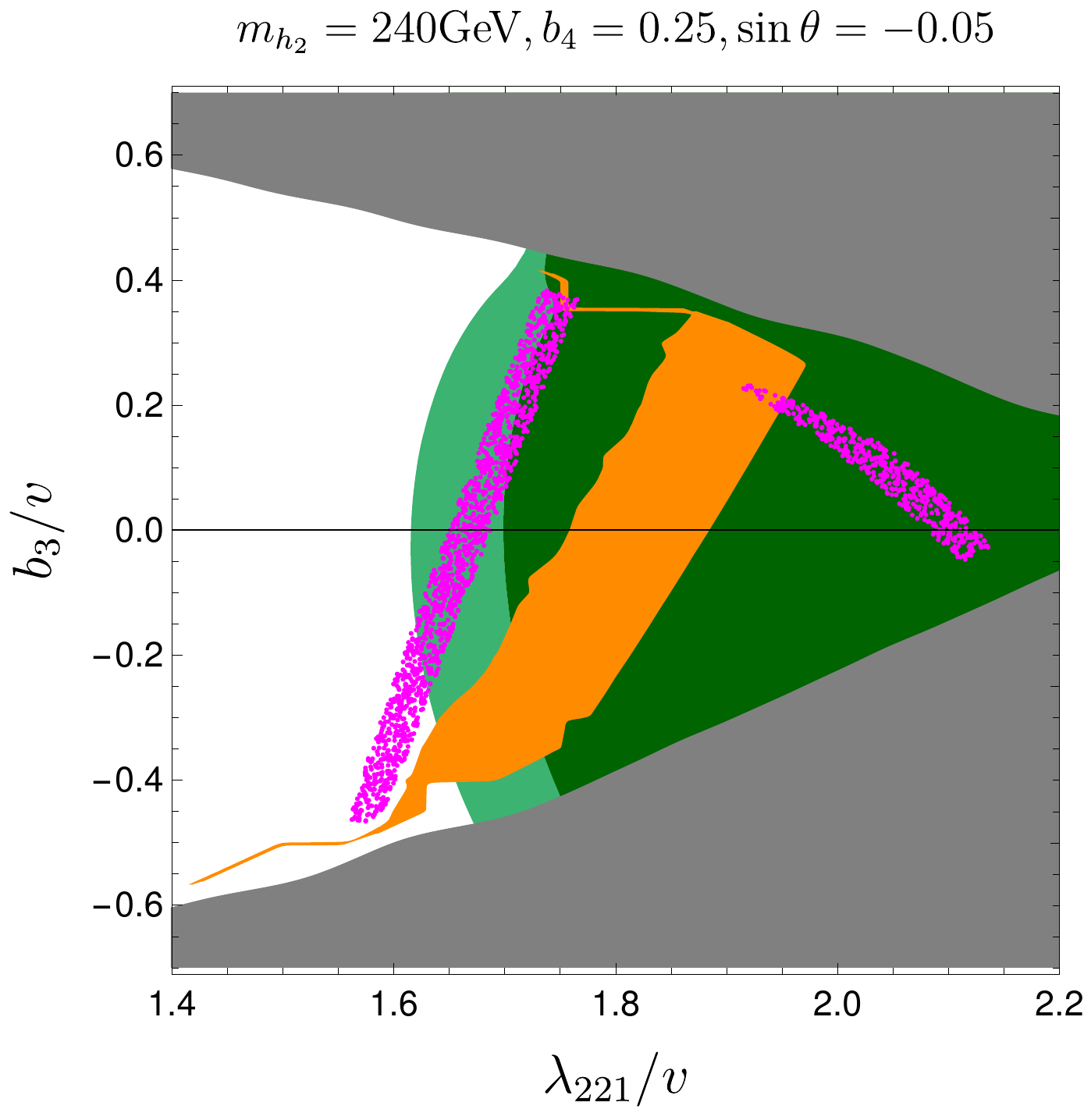}
  \quad
  \includegraphics[width=0.44\textwidth]{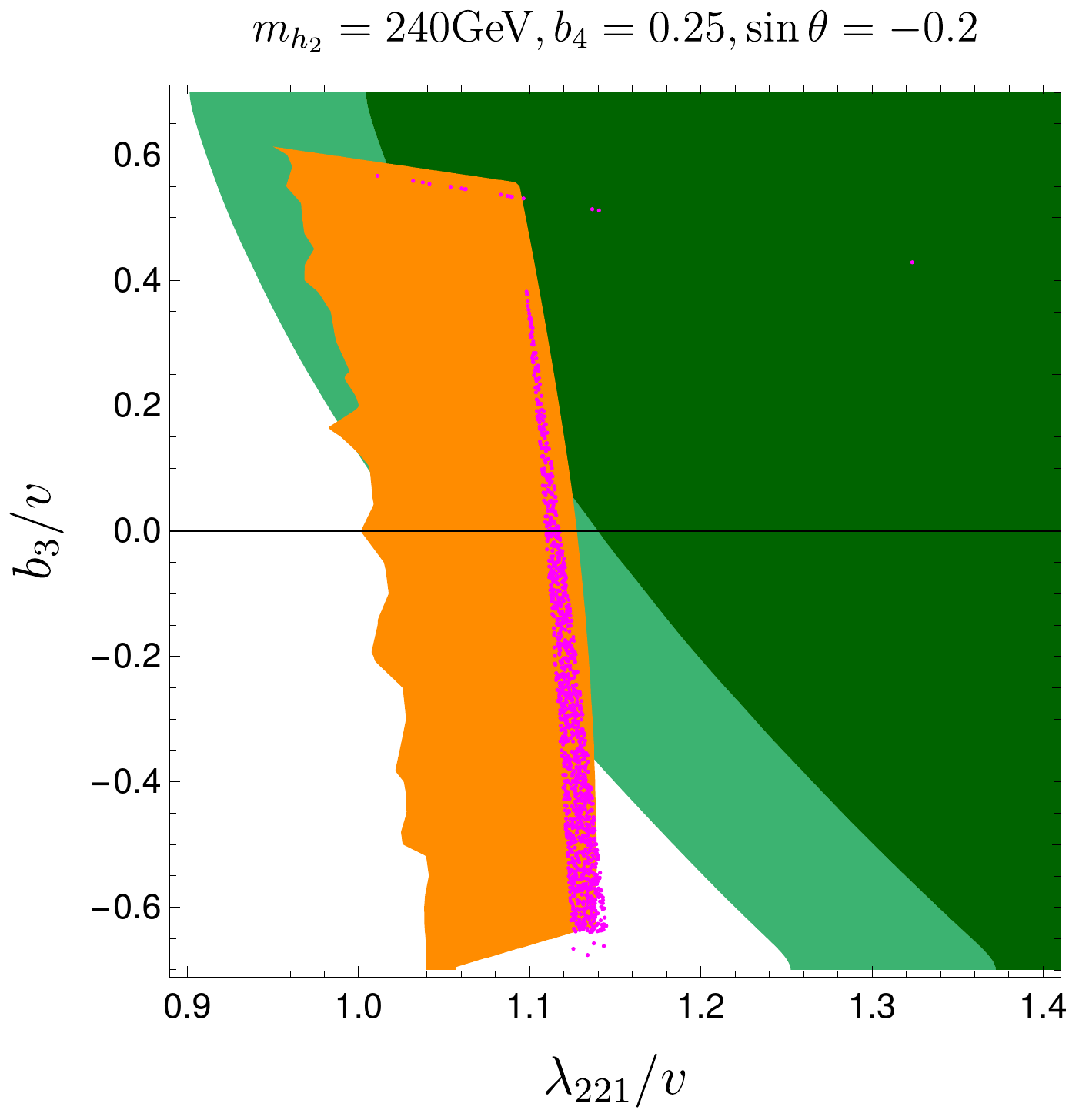}
  \\
  \includegraphics[width=0.44\textwidth]{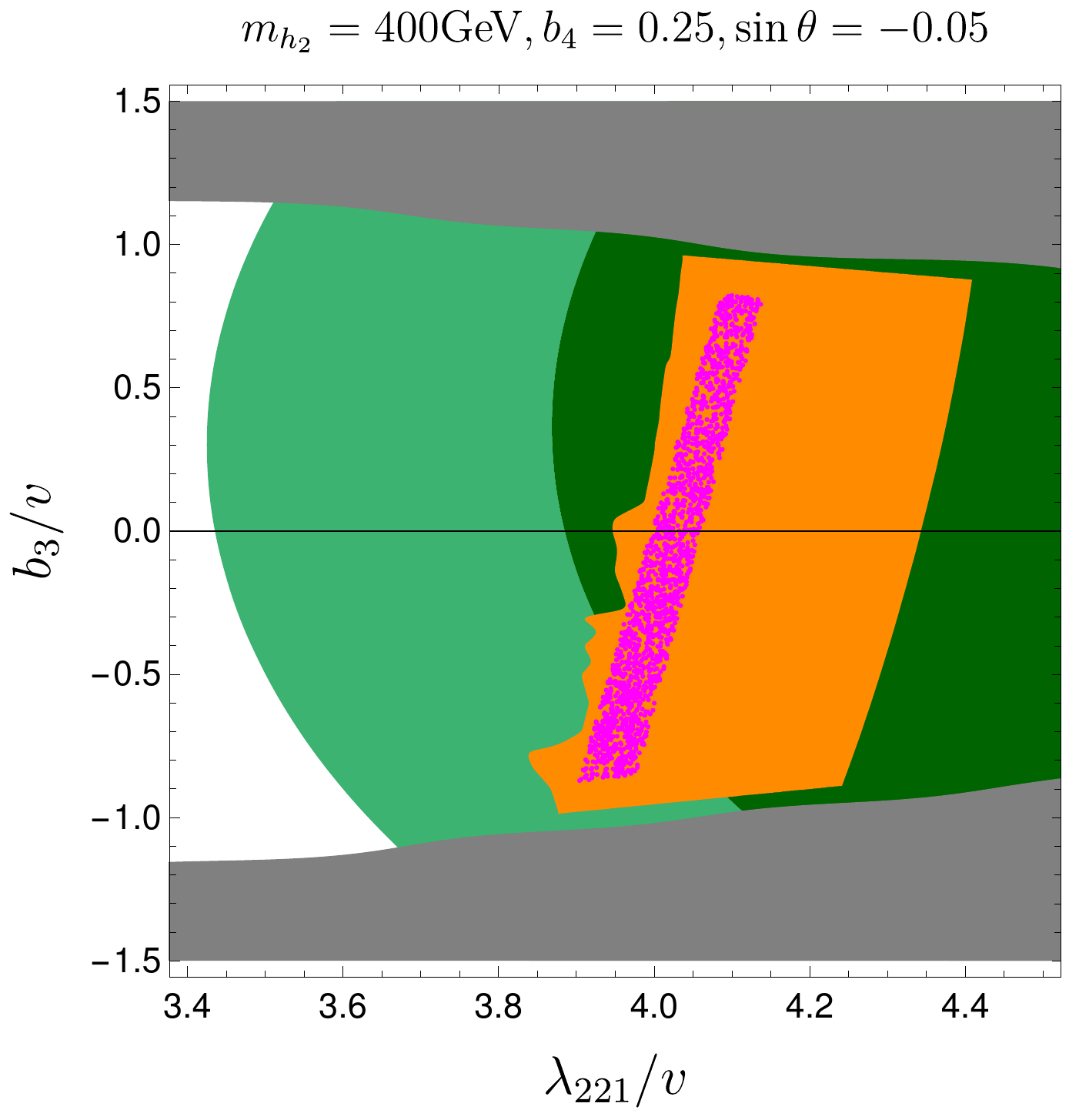}
  \quad
  \includegraphics[width=0.44\textwidth]{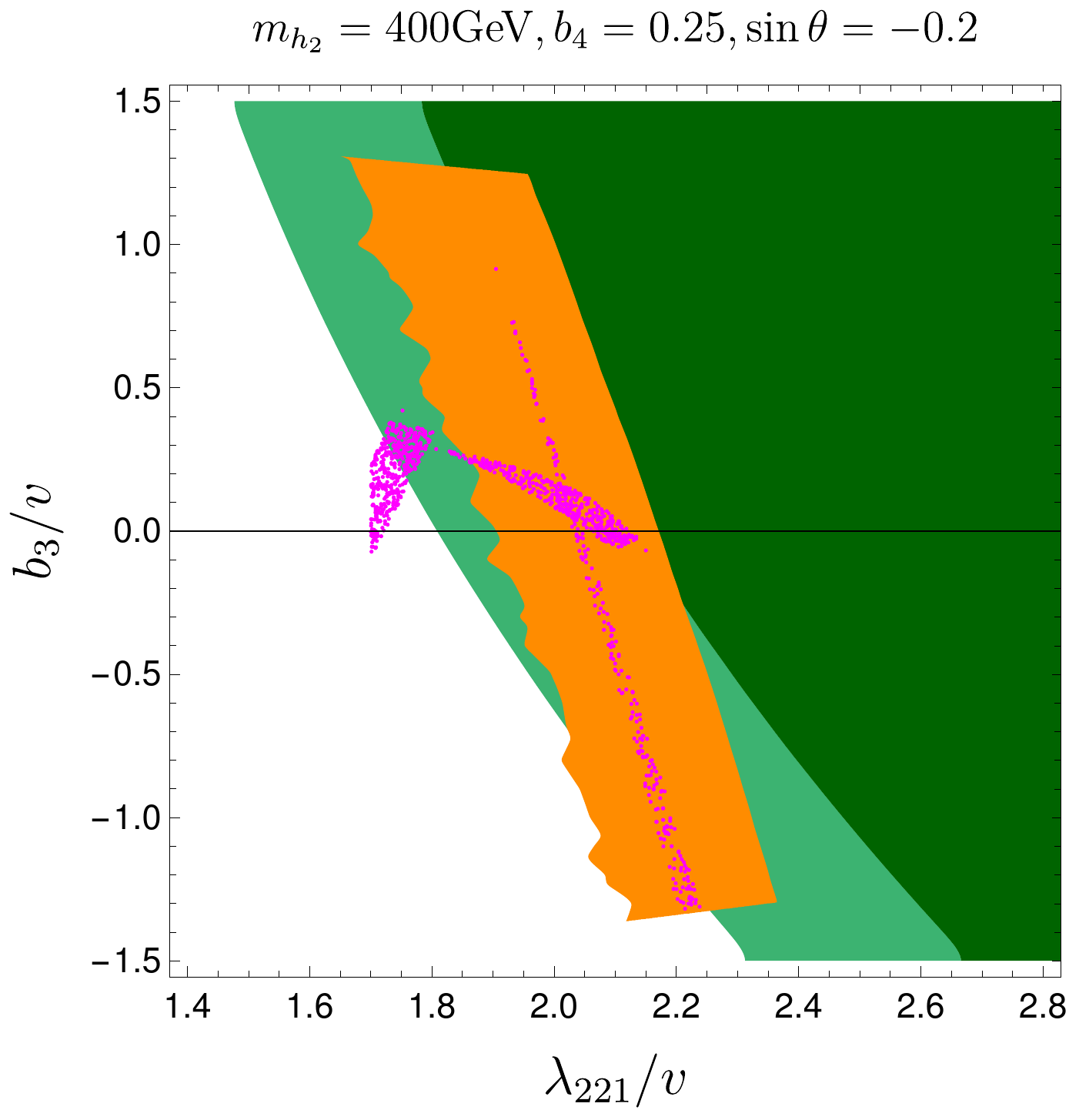}
\caption{
\label{fig:compare}
{Comparison with the results from the non-perturbative study in Ref.~\cite{Gould:2019qek}. Here $b_3$ is in 
the tadpole basis where a shifting of the $s$ field leads to $v_s=0$ and the appearance of a tadpole term in Eq.~\ref{eq:v}. 
The convertion between these two sets of parameters can be found in the appendix of~\cite{Alves:2018jsw}.
$\lambda_{221}$ is the coupling of $h_2 h_2 h_1$: $V \in i \lambda_{221} h_2^2 h_1/2$. Aside from $b_3$, all the other parameters shown here
are basis invariant.
The minus sign of $\theta$ is due to a different definition of the mixing angle from~\cite{Gould:2019qek} (compare Eq.~\ref{eq:mixing} with 
Eq.A2-A3 in~\cite{Gould:2019qek}).
These figures correspond to the four plots in Fig.5 of~\cite{Gould:2019qek}, where the light-green region gives
a first order phase transition from their non-perturbative study, the dark-green region indicates where higher dimensional operators 
is needed in the effective theory framework, the yellow one gives the corresponding region
from a full one-loop perturbative analysis, and the gray region is excluded by vacuum stability requirement. 
All these regions are taken from Ref.~\cite{Gould:2019qek}.
Overlaid magenta points on these figures are from our study where a gauge-independent high temperature expansion is 
adopted when calculating the finite temperature effective potential.}
}
\end{figure}

\section{\label{sec:gw}Gravitational Waves}

During the EWPT, a stochastic background of GW can be generated. In contrast to GW from a binary system, the amplitude of the stochastic background is a random variable, which is unpolarized, isotropic, and follows a Gaussian distribution~\cite{Allen:1997ad}. Therefore, it is characterized by the two-point correlation function and is proportional to the power spectral density, $\Omega_{\text{GW}}(f)$. Due to its stochastic origin, the detection method is also different. With one detector, this signal would behave as another source of noise, making its identification difficult. Thus, the detection of this kind of GW depends on cross-correlating the outputs from two or more detectors (for recent reviews on cosmological sources of stochastic GW,  see Ref.~\cite{Caprini:2018mtu} and for detection methods, see Ref.~\cite{Romano:2016dpx}). 

Given a particle physics model, the starting point of calculating the GW is the finite temperature effective potential. 
A standard perturbative calculation of the effective potential requires including the Coleman-Weinberg term~\cite{Coleman:1973jx}, the finite 
temperature corrections~\cite{Quiros:1999jp} and the daisy resummation~\cite{Parwani:1991gq,Gross:1980br}. However, out of concerns of the 
gauge dependence of the resulting effective potential (see e.g.,~\cite{Patel:2011th} for a discussion), we choose to use the high temperature approximation which is gauge-independent, 
following previous analyses~\cite{Profumo:2007wc,Kotwal:2016tex,Li:2019tfd,Alves:2018oct,Alves:2018jsw}. 
As the barrier in this model comes from the tree level cubic terms, the above approximation is better justified~\footnote{
Eventually, one might resort to the non-perturbative lattice simulation of the 3-dimensional effective theory, which is dimensionally reduced from the
4-dimensional full theory~\cite{Moore:2000jw}. This method is free of the gauge dependence issue and the infrared problem~\cite{Linde:1980ts}, but computationally expensive. For the particular model of xSM, the dimensional reduction was performed in~\cite{Brauner:2016fla}, with the resulting
EWPT and GW recently studied in~\cite{Gould:2019qek} based on an earlier lattice simulation result~\cite{Moore:2000jw}.
}.
From this effective potential, a set of portal parameters,
\begin{eqnarray}
\centering
T_n, \quad \alpha, \quad \beta/H_n, \quad v_w, \quad \kappa_v, \quad \kappa_{\text{turb}},
\label{eq:ptparams}
\end{eqnarray}
which characterize the dynamics of the EWPT may be calculated. Here, $T_n$ is the nucleation temperature and quantifies the time epoch when the bubbles are nucleated with
a large probability; $\alpha$ is the 
energy density released from the EWPT normalized by the total radiation energy density at $T_n$; $\beta/H_n$ describes, approximately, the
inverse time duration of the EWPT and also serves as a length scale for GW spectra, such as the peak frequency; $v_w$ is the bubble 
wall velocity; $\kappa_v$ is the fraction of released energy transferred into the kinetic energy of the plasma; and $\kappa_{\text{turb}}$ is
the fraction of energy going to the Magneto-Hydrodynamic-Turbulence (MHD).
It is this set of portal parameters that determine the GW spectra. 

\begin{figure}[t]
\centering
\includegraphics[width=0.6\textwidth]{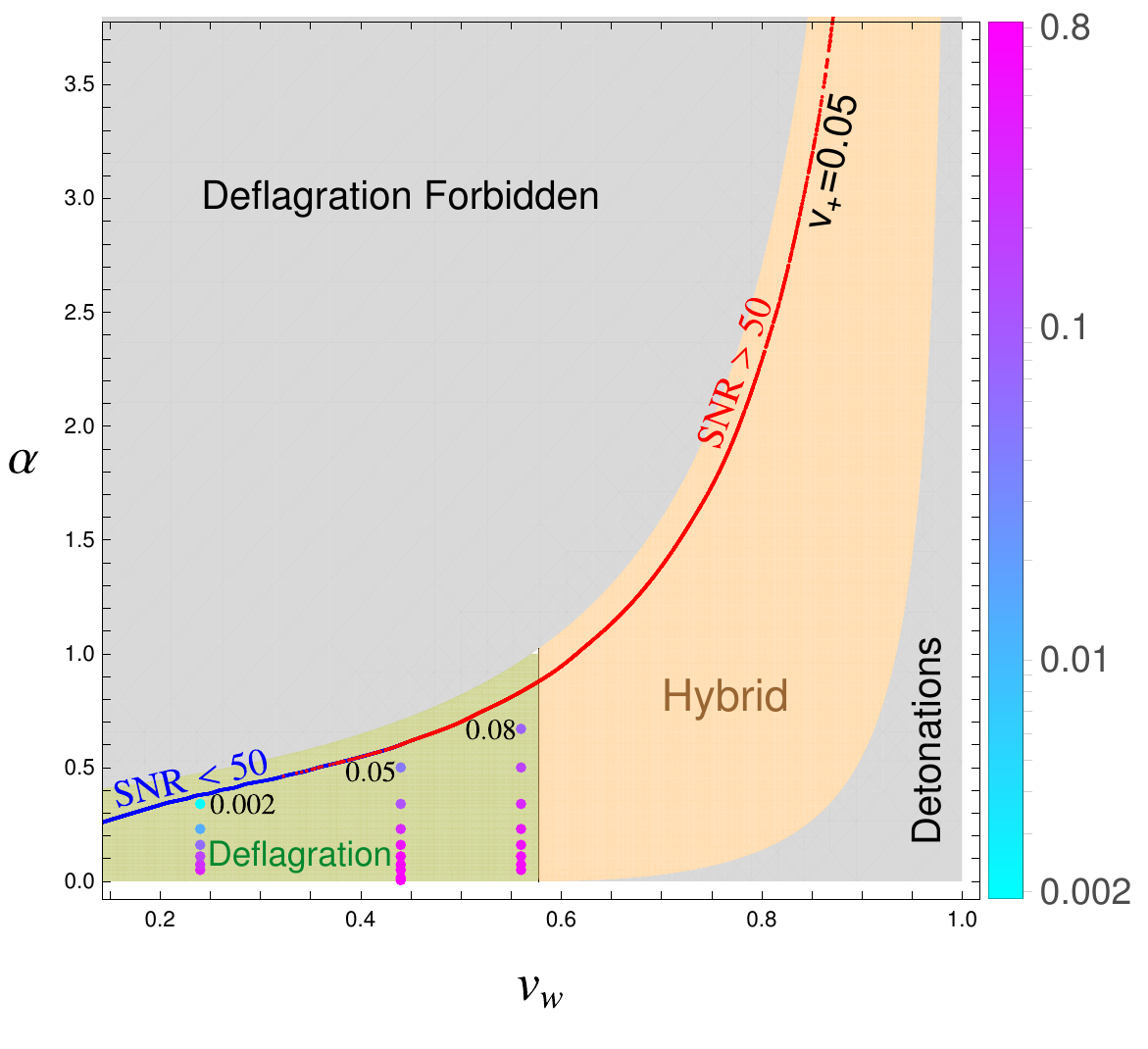}
\caption{
\label{fig:vwalpha}
This figure shows the normalized energy released from the EWPT $\alpha$ versus the bubble wall velocity $v_w$.
The gray region in the  top left is theoretically inaccessible and the gray region in
the  bottom right gives a detonation fluid profile which fails EWBG. The green region 
gives deflagration mode and the brown region gives hybrid mode. All ($v_w$, $\alpha$) on the
curve labelled $v_+=0.05$ give a plasma relative velocity 0.05 heading towards the wall from outside the bubble. All the scanned points fall on this curve, with the red (blue) points giving SNR larger (smaller) than 50, which are calculated without including the suppression factor (see the text for the details of the suppression factor). The dots denote the pairs of ($v_w$, $\alpha$) for which
numerical simulations of GW production are performed in Ref.~\cite{Cutting:2019zws}, 
with the color of each point showing the value of the suppression factor of
the GW signal as can be read from the color legend. The number close to the top-most point
in each of the three columns of points denotes the corresponding suppression factor for that
point.
}
\end{figure}

{
Before going to the details, we pause here and make a comparison with the recent result in Ref.~\cite{Gould:2019qek}, which is obtained based on
a dimensionally reduced 3-dimensional effective theory and previous lattice simulations. Ideally, their results are reliable when the second scalar, $s$ here, plays no dynamical role and is heavy so that it can be integrated out. However, a rough agreement has been found with the result from a perturbative determination of the effective potential, which includes the full one-loop terms and thus is gauge dependent.
It would be interesting to compare their findings with ours as a cross-check of our implementations. Therefore we digitize their Fig.5, as shown in 
Fig.~\ref{fig:compare}, and overlay our corresponding result in these plots(see caption for more details). 
In~\cite{Gould:2019qek}, the light-green and yellow regions are obtained by requiring $v_h(T_c)/T_c \in [0.3,0.6]$. 
Our magenta points, except for in the bottom-right panel, are also chosen with this condition, and they further give successful bubble nucleations, 
i.e., the nucleation temperature $T_n$ can be obtained. For the bottom-right panel, we have to slightly increase the upper boundary, such as by choosing $v_h(T_c)/T_c \in [0.3,0.8]$, to reveal the band with the same shape as the yellow one.
We can see the regions capable of generating a first-order phase transition by these three different calculations are located at roughly the same place, though they differ slightly in the fine structures. This justifies our approach of calculating
the effective potential, which maintains gauge invariance. While one may doubt the high-temperature approximation, as for some regions
of the parameter space the finite temperature vevs at the phase transition are not very smaller than the temperature, the tree level barrier rather
than quantum corrections as the main cause of the phase transition guarantees that the result is not sensitive to this approximation.
Due to the highly expensive and thus so-far limited set of lattice simulations, there is currently only limited coverage of the parameter space for the non-perturbative study, which makes a full comparison with ours impossible as of now. However, should future lattice simulations become 
available for the parameter space where the additional scalar plays a dynamical role, it is essential to continue this line of comparisons.
}

Now we go to the GW calculations. Since we have given a detailed description of the calculations of these parameters in the xSM in Ref.~\cite{Alves:2018jsw}, here we present only a summary of the key physical highlights adopted in our calculations, and the new features of the current study.

\begin{itemize}
\item {\bf $v_w$ and Hydrodynamics --} A stronger GW production  usually requires a larger $v_w$, while a successful EWBG needs a small, generally subsonic $v_w$ (for EWBG calculations, see e.g.,~\cite{John:2000zq,Cirigliano:2006dg,Chung:2009qs,Chao:2014dpa,Guo:2016ixx}). Hence, a large $v_w$, which can source detectable GW signals, is detrimental to the process of generating the observed baryon asymmetry of the universe. The solution to this tension adopted in this work follows Ref.~\cite{No:2011fi} and hinges on the recognition that $v_w$ may not be the quantity that enters EWBG calculations, due to the non-trivial plasma velocity profile surrounding the bubble wall. This can be further understood by noting that there exist three fluid velocity profiles for a single bubble in the phase transition:  deflagration, supersonic deflagration (also known as hybrid), and detonation (see Ref.~\cite{Espinosa:2010hh} for a recent combined analysis). In the cases of deflagration and supersonic deflagration, the fluid has a non-zero velocity outside the bubble wall. From the perspective of the bubble wall frame, the fluid would head towards the wall with a velocity ($\equiv v_+$) that is smaller than $v_w$. While the justification of this argument still needs a combined analysis of both macroscopic bubble behavior and microscopic particle transport dynamics, we assume tentatively that it is true in our series of papers on this subject~\cite{Alves:2018oct,Alves:2018jsw,Bian:2017wfv,Bian:2019zpn}. This implies that in the plane of ($v_w$, $\alpha$), as shown here in Fig.~\ref{fig:vwalpha}, only certain regions are compatible with EWBG. In this plot, the green and brown regions denote the deflagration and supersonic deflagration (hybrid) modes of the plasma, while the gray regions are either
theoretically not allowed or not compatible with EWBG. The curve denotes the location of all the scanned points in this study, corresponding to $v_+ = 0.05$, the usually adopted benchmark for EWBG calculations. 

\item {\bf Efficiency factors $\kappa_v$ and $\kappa_{\text{turb}}$ --} 
The energy fraction transferred to the fluid kinetic energy $\kappa_v$ is an important parameter as it directly determines the strength of the GW signal coming from the sound waves. In the past, it has been calculated as a function of ($v_w$, $\alpha$) by first solving the fluid velocity profile for a single bubble and then calculating the kinetic energy from the energy-momentum tensor~\cite{Espinosa:2010hh,Chao:2017ilw}. The resulting value agrees well with values inferred from numerical simulations
for relatively weak phase transition, defined such that $\alpha \ll 1$. This agreement had motivated the generalization of the obtained GW formula for arbitrary values of $\alpha$. 

This naive generalization, however, has been proven to be wrong by a very recent numerical simulation result~\cite{Cutting:2019zws}. In this study, several sets of simulations have been performed for strong phase transitions, defined by the authors as $\alpha \sim 1$. Their study shows a significant deficit in the produced GW signals when the plasma is of the deflagration mode while for detonations, it is less affected. There is currently no simulation result available for supersonic deflagrations.

The suppression for the deflagration plasma is found to be due to the reduction of $\kappa_v$. The physical reason is that the formation of reheated droplets of the unbroken phase slows down the bubble expansion. This disagreement reveals a missing piece in the understanding of the fluid dynamics and requires a more accurate theoretical modeling in the future, which, however, is beyond the scope of our work. Therefore, we use the results from the limited set of numerical simulations given in
\blue{the Table I of} Ref.~\cite{Cutting:2019zws} and quantify their effects for generic choices of $(v_w, \alpha)$~\footnote{The quantities characterizing our EWPT include other parameters such as $T_c$ and $T_n$, whose values are different from what was used in the numerical simulations in~\cite{Cutting:2019zws}. We assume these additional parameters to be of sub-dominant effect compared with ($v_w$, $\alpha$).}.

    We denote the reduction of GW amplitude as $\delta$ and show its values for the set of simulations performed in Ref.~\cite{Cutting:2019zws} in the ($v_w$, $\alpha$) plot in Fig.~\ref{fig:vwalpha}. Here the three columns of dots denote the values of $\delta$ for the corresponding values of ($v_w$, $\alpha$), with the color characterizing its value, which can be read from the legend to the right of the plot. The number associated with the point at the top of each column explicitly denotes the corresponding value of $\delta$. 
We can see that there is a generic suppression for all cases, and $\delta$ sharply drops for larger values of $\alpha$. 
But as $v_w$ is increased, $\delta$ is less reduced. Since results covering the whole plane of 
($v_w$, $\alpha$) are currently unavailable, we choose $\delta=0.01$ as a conservative estimate of the GW signals~\footnote{This modified bubble growth picture would also affect the $v_w-\alpha$ curve in Fig.~\ref{fig:vwalpha} and needs further study.}.

For the energy fraction going to the MHD, we note that the current simulations are not long enough for MHD to fully develop. It is found that $\kappa_{\text{turb}} \approx (0.05 \sim 0.1) \kappa_v$~\cite{Hindmarsh:2015qta}. We choose  $\kappa_{\text{turb}}=0.1 \kappa_v$ in our work.

\begin{figure}
\centering
\includegraphics[width=0.4\textwidth]{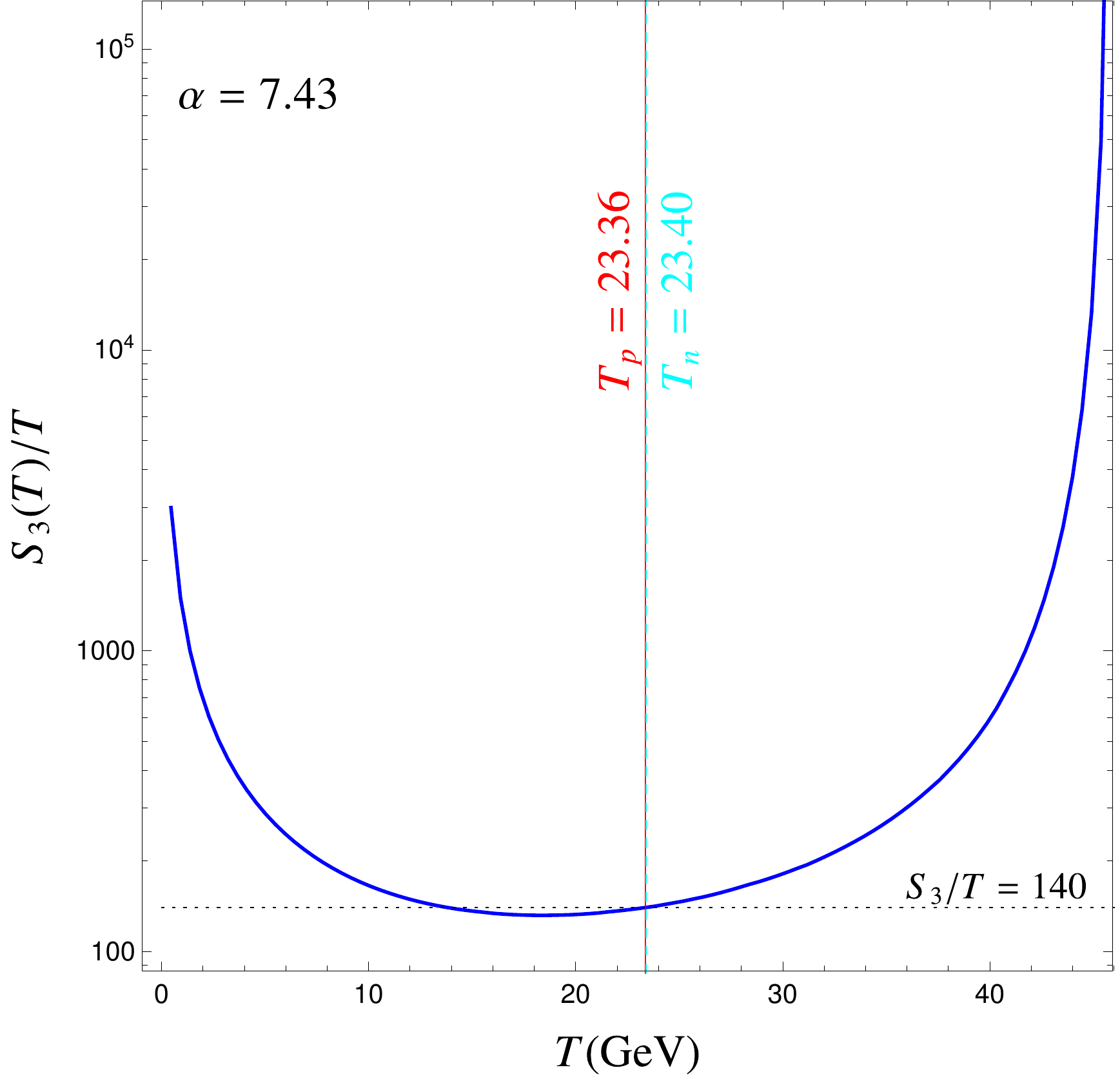}
\quad
\includegraphics[width=0.4\textwidth]{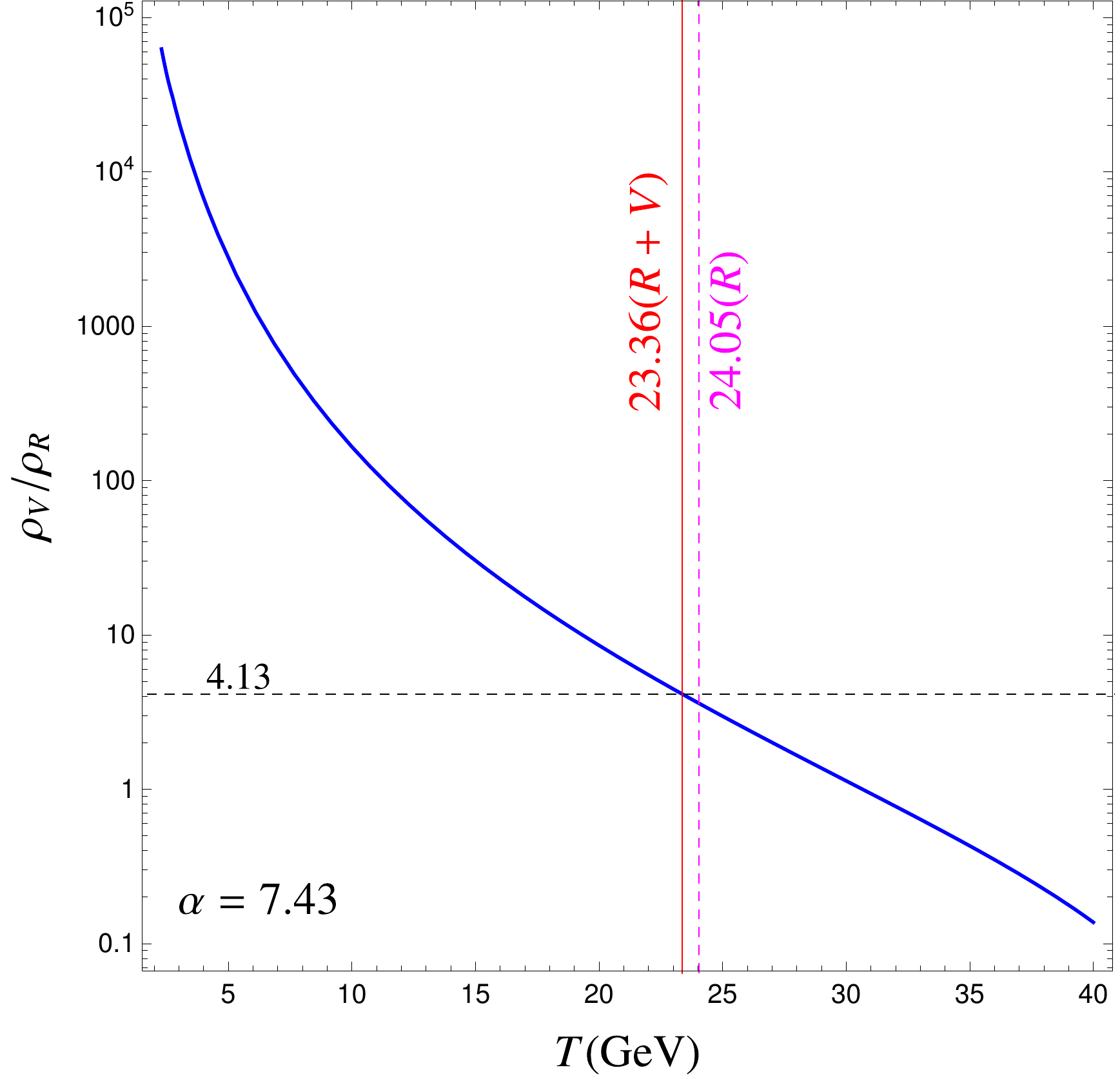}
\caption{\label{fig:perco}
The details of a super strong EWPT with $\alpha=7.43$. The left panel shows the 3-dimensional Euclidean action of the critical bubble divided by $T$ as 
$T$ drops. $T_n$ (cyan dashed vertical line) is obtained at $S_3/T=140$, corresponding to the intersection between the blue curve and the horizontal dotted line. 
The percolation temperature $T_p$ is also shown with the vertical red line. The right panels shows the $\rho_V \equiv \Delta V$ normalized by the total 
radiation energy density $\rho_R$ for this benchmark. Here the red vertical line corresponds to $T_p$ and the slightly higher (magenta dashed line) is obtained
by simply neglecting $\Delta V$ in the Hubble expansion rate. Here ``R'' and ``V'' denote the 
contribution of radiation and vacuum energy density
respectively.
}
\end{figure}
\item {\bf Supercooled EWPT --}
Strong GW signals generally require more energy released from the EWPT, corresponding to more supercooled phase transitions. This is primarily so due to the previously mentioned reduction in GW production, which makes 
generating a detectable GW signal more difficult and requires even more supercooled EWPT.

The environment within which the bubbles are nucleated may then be vacuum energy dominated rather than dominated by radiation. In this case, the space outside the bubbles will inflate, and it may happen
that the bubbles never meet each other. The universe would be trapped in a metastable vacuum. Therefore
there exists a maximal phase transition strength, above which the EWPT is not feasible~\cite{Ellis:2018mja}. 
To guarantee that EWPT does, in fact, complete, one must ensure the stronger condition that the physical volume (as opposed to the comoving volume) of the unbroken space is shrinking.

The temperature at which the above condition is imposed is chosen to be the percolation temperature in Ref.~\cite{Ellis:2018mja}.
Here the percolation temperature ($\equiv T_p$) is defined such that approximately $30\%$ of the spatial volume is in the symmetry broken phase. We note that $T_p$ is typically slightly less than $T_n$. In the xSM, we found that for most of the points we used in Ref.~\cite{Alves:2018jsw},
$T_n$ is less than $T_p$ by $2\%$, and the stronger percolation criterion can be satisfied. 

As an example, we show the details of a super-strong EWPT with $\alpha = 7.43$ in Fig.~\ref{fig:perco}. The left panel shows 
$S_3(T)/T$ as a function of $T$ and the positions of $T_n$ (cyan dashed) and $T_p$ (red), which are barely distinguishable from 
each other. The right panel shows $\Delta V$, the difference between the potential energy density at the false and true vacua, normalized by the radiation energy density $\rho_R$. Here aside from $T_p$ which was solved using the full Hubble expansion rate, we also present a higher temperature obtained by neglecting $\Delta V$ in $H$, to show the impact on the obtained percolation temperature with and without the vacuum energy. So for this parameter choice, when $\alpha$ is large, the two temperatures do not differ substantially. We numerically checked that for this benchmark the physical volume of the unbroken phase does shrink at $T_p$. 

\item {\bf One or Two-Step EWPT --} The EWPT in this model can proceed in two different patterns. The first pattern is a direct transition into the EW broken phase, while the second involves two stages, with the first one giving the 
extra singlet a non-zero vev followed by a second transition to the EW broken phase~\cite{Patel:2012pi,Ramsey-Musolf:2017tgh,Chao:2017vrq,Alves:2018jsw}. Most of the parameter space in this model involves a single step transition, while regions yielding two-stage transitions reside in a  totally different part of parameter space~\cite{Alves:2018jsw}. We thus focus on points sampled only from the parameter space of one-step EWPT.

\end{itemize}

The strategy described above is followed in calculating the set of parameters in Eq.~\ref{eq:ptparams}. The GW spectra can be obtained by plugging these parameters 
into a set of analytical formulae, obtained by fitting to results from numerical simulations of different GW production mechanisms.
It has been realized in recent years~\cite{Hindmarsh:2013xza} that the dominant contribution comes from sound waves, although significant advances have also been made in both analytical modeling and numerical simulations of pure bubble collision contributions~\cite{Jinno:2017fby,Jinno:2017ixd,Jinno:2016vai,Cutting:2018tjt}. 
By evolving the scalar-field and fluid model on a 3-dimensional lattice, 
the gravitational wave energy density spectrum can be extracted~\cite{Hindmarsh:2015qta}:
\begin{eqnarray}
\Omega_{\textrm{sw}}h^{2}=2.65\times10^{-6}
\delta\times
  \left( \frac{H_{\ast}}{\beta}\right) \left(\frac{\kappa_{v} \alpha}{1+\alpha} \right)^{2} 
\left( \frac{100}{g_{\ast}}\right)^{1/3} 
\times v_{w} \left(\frac{f}{f_{\text{sw}}} \right)^{3} \left( \frac{7}{4+3(f/f_{\textrm{sw}})^{2}} \right) ^{7/2} \ .
\label{eq:soundwaves}
\end{eqnarray}
Here $g_{\ast}$ is the number of relativistic degrees of freedom and $H_{\ast}$ is the Hubble parameter at $T_{\ast}$, both evaluated at a time when the phase transition 
has just completed. 
Moreover, $f_{\text{sw}}$ is the present day peak frequency of this spectrum:
 \begin{equation}
f_{\textrm{sw}}=1.9\times10^{-5}\frac{1}{v_{w}}\left(\frac{\beta}{H_{\ast}} \right) \left( \frac{T_{\ast}}{100\textrm{GeV}} \right) \left( \frac{g_{\ast}}{100}\right)^{1/6} \textrm{Hz} .
\end{equation}
We have inserted an ad hoc factor $\delta$ in this formula, to take into account the reduction of the gravitational waves, as discussed
earlier. We choose a conservative value of $\delta=0.01$, to be consistent with results of the sets of simulations performed in Ref.~\cite{Cutting:2019zws}.

In addition, the fully ionized plasma can result in the formation of magnetohydrodynamic (MHD) turbulence. This can be calculated analytically
by assuming a proper power spectrum for the turbulence as well as the primordial magnetic field~\cite{Kosowsky:2001xp,Caprini:2009yp,Gogoberidze:2007an,Niksa:2018ofa} or numerically by evolving
the magnetic field governed by the MHD differential equations and also coupled to gravity~\cite{Pol:2019yex,Brandenburg:2017neh}. 
We use the result presented in~\cite{Caprini:2009yp,Binetruy:2012ze},
\begin{eqnarray}
\Omega_{\textrm{turb}}h^{2}=3.35\times10^{-4}\left( \frac{H_{\ast}}{\beta}\right) \left(\frac{\kappa_{\text{turb}} 
\alpha}{1+\alpha} \right)^{3/2} \left( \frac{100}{g_{\ast}}\right)^{1/3}  
v_{w}  \frac{(f/f_{\textrm{turb}})^{3}}{[1+(f/f_{\textrm{turb}})]^{11/3}(1+8\pi f/h_{\ast})} ,
\label{eq:mhd}
\end{eqnarray}
with $f_{\text{turb}}$ being the peak frequency:
\begin{equation}
f_{\textrm{turb}}=2.7\times10^{-5}\frac{1}{v_{w}}\left(\frac{\beta}{H_{\ast}} \right) \left( \frac{T_{\ast}}{100\textrm{GeV}} \right) \left( \frac{g_{\ast}}{100}\right)^{1/6} \textrm{Hz} .
\end{equation}
With the energy density spectrum obtained and used as a matched filter, the outputs from a pair of gravitational wave detectors can be cross correlated to 
search for the signal. The noise of each detector drops out of the ensemble average (which is equivalent to a time average) of this cross correlation, leaving the desired signals~\footnote{
Noises common to both detectors need also be subtracted.}. The detectability of the gravitational waves is then quantified by the signal-to-noise ratio (SNR)~\cite{Caprini:2015zlo}: 
\begin{eqnarray}
  \text{SNR} = \sqrt{\mathcal{T} \int_{f_{\text{min}}}^{f_{\text{max}}} df 
    \left[
      \frac{h^2 \Omega_{\text{GW}}(f)}{h^2 \Omega_{\text{exp}}(f)} 
  \right]^2} ,
\end{eqnarray}
where $\mathcal{T}$ is the duration of the time series of the detector ouput in years and 
$\Omega_{\text{exp}}(f)$ is a similarly defined detector power spectral density for this pair of 
interferometers. Several such detectors have been proposed, including the Laser Interferometer 
Space Antenna (LISA)~\cite{Audley:2017drz}, the 
Big Bang Observer (BBO), the DECi-hertz Interferometer Gravitational wave 
Observatory (DECIGO)~\cite{Yagi:2011wg}, Taiji~\cite{Gong:2014mca} and Tianqin~\cite{Luo:2015ght}. 
In this study, we will focus on LISA as a benchmark detector and assume $\mathcal{T}=5$.
It was suggested in Ref.~\cite{Caprini:2015zlo} that for a four-link LISA configuration, the threshold
for detection is $\text{SNR}=50$ and for a six-link configuration, an SNR as low as 10 can be used.

\section{\label{sec:4b}Collider Analysis}

Having discussed the production of GW, in this section, we present our collider analysis. Our focus is on double Higgs production at the 13~TeV HL-LHC. We access this production mode via the $pp\rightarrow h_1(b\bar{b})h_1(b\bar{b})$ channel, demanding four resolved $b$-tagged jets. The main backgrounds for this search are the QCD multi-jet and $\bar{t}t$. 

 In previous papers, we have studied the GW complementarity in the $h_2 \to h_1h_1\to \bar{b}b\gamma \gamma$~\cite{Alves:2018oct} and $h_2 \to VV \, (V=W,Z) $~\cite{Alves:2018jsw} channels. In Ref.~\cite{Alves:2018oct}, some representative xSM benchmark points with favorably large SNR were investigated and the potential of the LHC to discover a double Higgs signal in $\bar{b}b\gamma \gamma$ using machine learning techniques were explored.  On the other hand, the analysis of Ref.~\cite{Alves:2018jsw} established the complementary role of the LHC searches of $h_2$ in diboson channel by merely extrapolating recent ATLAS and CMS results to 3 ab$^{-1}$. The $4b$ channel investigated in this work complements the results of those work for two reasons. First, $h_1h_1$ and $WW,ZZ$ are the main decay modes of $h_2$ in the considered parameter space. 
 Second, $b\bar{b}b\bar{b}$ is the double Higgs decay mode with the largest branching ratio.

 In order to constrain the xSM in the GW context, we performed a shape analysis of the $4b$ mass distribution~\cite{Goncalves:2018yva,Biekotter:2018jzu}. The signal samples are generated with {\sc MadGraph5}~\cite{Alwall:2014hca}. Hadronization and underlying event effects are accounted for with {\sc Pythia8}~\cite{Sjostrand:2014zea} and detector effects are simulated with {\sc Delphes3} package~\cite{deFavereau:2013fsa}. Higher order corrections are included with a next-to-next-to-leading order QCD $K$-factor~\cite{Borowka:2016ypz,deFlorian:2016uhr,deFlorian:2013uza} for the non-resonant component of the signal. The resonant  contribution to the the signal is not expected to receive a very different QCD correction~\cite{Catani:2003zt,Dawson:2015haa} so we keep the same K-factor  to all contributions.

In our analysis, we closely follow the ATLAS study, Ref.~\cite{Aaboud:2018knk}. ATLAS models the background with a data-driven approach.  This is a more reliable procedure as the multi-jet component displays very large  QCD corrections that are challenging to realistically account for in a Monte Carlo simulation. Hence, we use the  backgrounds from ATLAS in our analysis. We validate our signal simulations using the SM double Higgs production presented by ATLAS~\cite{Aaboud:2018knk}. We observe excellent agreement for the SM double Higgs mass distribution.

We start our analysis by requiring  four isolated $b$-tagged jets with $p_{Tj}>40$~GeV and ${|\eta_j|<2.5}$. Jets are defined with a cone radius of $0.4$ using the anti-$k_t$ jet algorithm implemented in {\sc FastJet}~\cite{Cacciari:2008gp,Cacciari:2011ma}. The $b$-tagging requirements use the working point with a $b$-tagging efficiency of  $70\%$  associated to a mistag rate of 15\% for $c$-quarks and 
0.2\% for light flavours. 
\begin{figure}[t]
\centering
\includegraphics[scale=0.62]{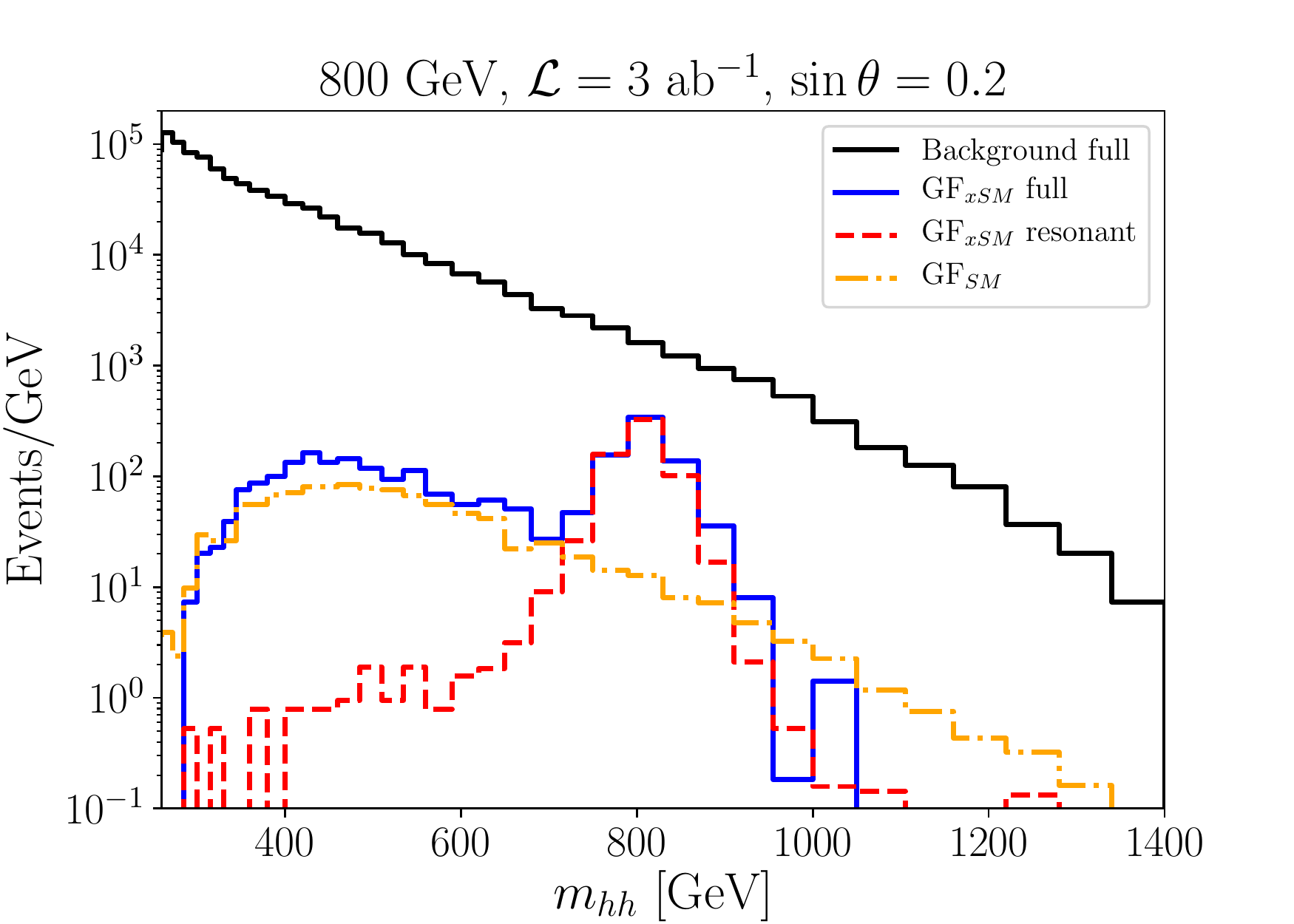}
\caption{$m_{hh}$ distribution for the GF components:  SM (yellow), xSM resonant (red), and xSM full that accounts for  the resonant and non-resonant contributions (blue). The total background component obtained from ATLAS via a data driven approach is also shown (black)~\cite{Aaboud:2018knk}. The signal sample is simulated with a Heavy Higgs mass $m_{h_2}=800$~GeV, $v_s=24.8$~GeV, $b_3/v_{EW}=-10$, $b_4=4.2$, and mixing $\sin\theta=0.2$. We consider a 13~TeV LHC with $\mathcal{L}=3$~ab$^{-1}$ of data.
\label{figure:mhh}}
\end{figure}

The four $b$-tagged jets reconstruct the two SM Higgs boson candidates. The pairings of jets into Higgs boson candidates are required
to satisfy:
\begin{alignat}{5}
\frac{360~\text{GeV}}{m_{4j}}-0.5 <& \Delta R_{jj, \text{lead}} <& \frac{653~\text{GeV}}{m_{4j}}+0.475 \,, \notag \\
\frac{235~\text{GeV}}{m_{4j}}\qquad\ <& \Delta R_{jj, \text{subl}} <& \frac{875~\text{GeV}}{m_{4j}}+0.35 \,,
\end{alignat}
where $\Delta R_{jj, \mathrm{lead}}$ ($\Delta R_{jj, \mathrm{subl}}$) is the angular distance between the jets that reconstruct the leading (sub-leading) SM Higgs 
boson candidate. In this first step of the analysis, the leading Higgs boson candidate is chosen to have the highest scalar sum of jet transverse momentum. To further reject the multi-jet background, we impose a pseudorapidity difference between the two Higgs candidates of $|\Delta\eta_{hh}|<1.5$. 

Mass-dependent selections on the Higgs boson candidates transverse momenta are imposed to further control the backgrounds
\begin{alignat}{5}
p_{Th}^{\text{lead}}>0.5 m_{4j}-103~\text{GeV}\,, \notag \\
p_{Th}^{\text{subl}}>0.33 m_{4j}-73~\text{GeV}\,.
\end{alignat}
The second step of the analysis chain requires a selection of $b$-tagged jet pairs associated to Higgs boson decays taking energy losses into account. The quantity 
\begin{equation}
    D_{hh}=\frac{m_{2j}^{lead}-\frac{120}{110}m_{2j}^{subl}}{\sqrt{1+\left(\frac{120}{110}\right)^2}}
\end{equation}
is computed and the pairing with the smallest $D_{hh}$ is chosen. These two pairs are then associated to $m_{2j}^{lead}$ and $m_{2j}^{subl}$ which are used to impose invariant mass selections around the SM Higgs boson mass for the leading and sub-leading Higgs boson candidates according to
\begin{equation}
X_{hh}=\sqrt{\left(\frac{m_{2j}^{\text{lead}}-120~\text{GeV}}{0.1m_{2j}^{\text{lead}}}\right)^2+
\left(\frac{m_{2j}^{\text{subl}}-110~\text{GeV}}{0.1m_{2j}^{\text{subl}}}\right)^2}<1.6\,.
\end{equation}

To further suppress the $\bar{t}{t}$ background, all possible combinations of three jets with one being $b$-tagged and  a constituent of the Higgs boson candidate are considered. The two light jets are considered as forming a hadronically decaying $W$ boson candidate. A measure of the compatibility with the top-quark candidate can be defined as 
\begin{equation}
X_{Wt}=\sqrt{\left(\frac{m_{W}-80~\text{GeV}}{0.1m_{W}}\right)^2+
\left(\frac{m_{t}-173~\text{GeV}}{0.1m_{t}}\right)^2}\,,
\end{equation}
where $m_t$ is the invariant mass of the three jet top candidate and $m_W$ is the two jet $W$ boson candidate. Events with 
the smallest $X_{Wt}<1.5$, from all possible three jet combinations,  are vetoed. To improve the signal $m_{4j}$ resolution, the four-momentum of each Higgs boson candidate is multiplied by the correction factor $m_{h_1}/m_{2j}$. This was found to improve the signal mass resolution by 30\%  with sub-leading impact on the background $m_{4j}$ distribution~\cite{Aaboud:2018knk}.

\begin{figure}[t]
\includegraphics[width=0.49\textwidth]{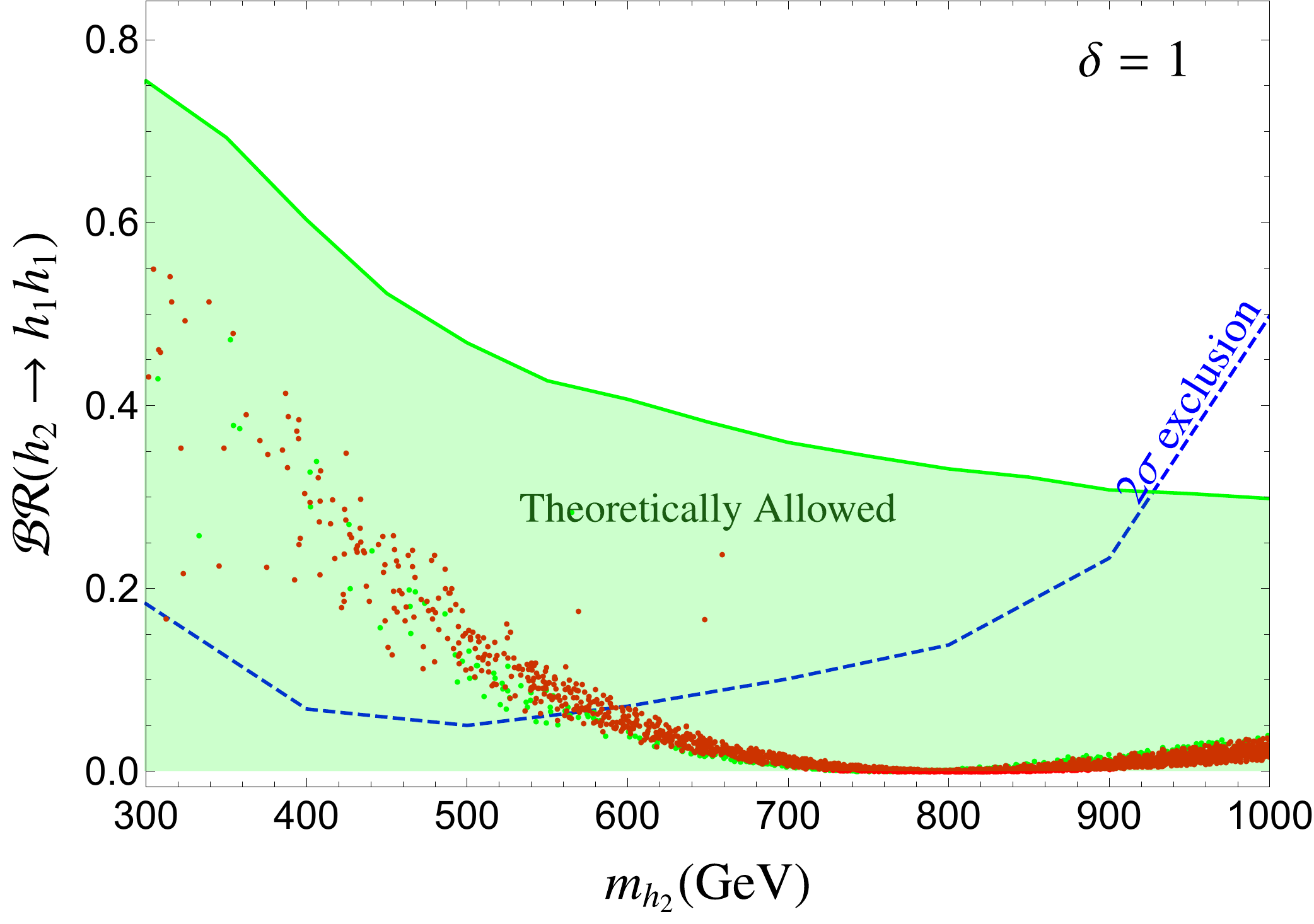}
\includegraphics[width=0.49\textwidth]{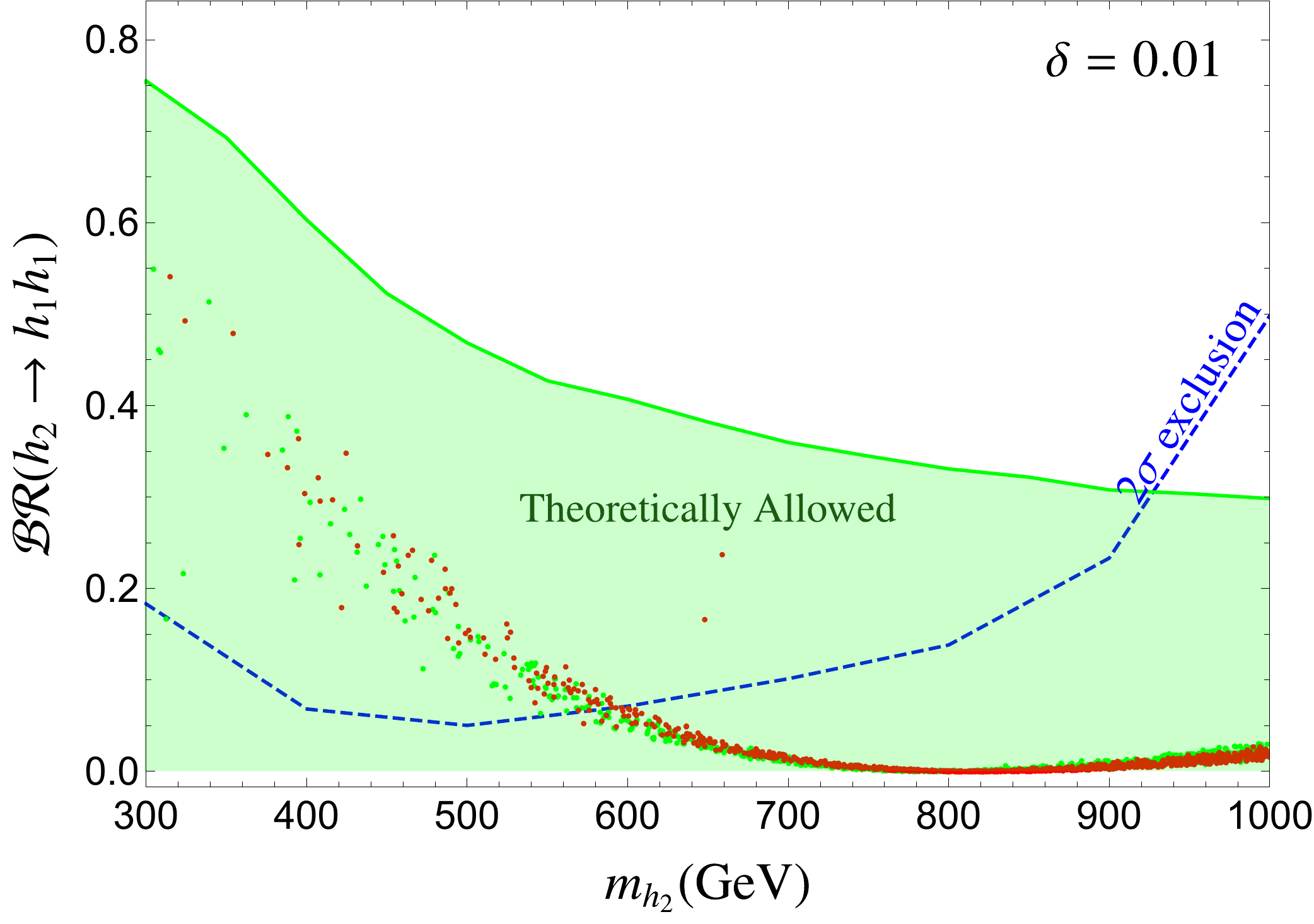}
\caption{$2\sigma$ exclusion bound (blue dashed line) for the $\mathcal{BR}(h_2\rightarrow h_1h_1)$ as a function of the 
heavy Higgs mass $m_{h_2}$. We assume the Higgs mixing $\sin\theta=0.2$ and consider the 13~TeV LHC with $\mathcal{L}=3$~ab$^{-1}$ of data.
The points where the resulting GW signal can be detected by LISA
are overlaid here, with the red color denoting those giving $\text{SNR}>50$, and green giving $10<\text{SNR}<50$. 
The left and right panels differ in that
the points in the right panels used a reduction factor of $\delta=0.01$ in calculating the GW spectra while those in the left panel
is obtained without this factor, i.e., $\delta=1$. 
Theoretical requirements on the potential such as the perturbative
unitarity, perturbativity, and vacuum stability at zero temperature generally impose
an upper bound on the branching ratio, corresponding to the green-colored region. 
\label{figure:bound}}
\end{figure}

In Fig.~\ref{figure:mhh}, we display the  double Higgs invariant mass $m_{hh}$ distribution for the Gluon Fusion (GF) contributions: SM, resonant xSM,  and full xSM (that accounts for  the resonant and non-resonant contributions). The total background component obtained from ATLAS with a data-driven approach is also shown~\cite{Aaboud:2018knk}.  The xSM distribution is illustrated with the parameter choice $m_{h_2}=800$~GeV and $\sin\theta=0.2$. We observe that the full xSM invariant mass $m_{hh}$ distribution display a significant contribution from the non-resonant GF terms. Hence, instead of only accounting for the resonant signal contribution, we describe the signal component as a deviation of the full xSM GF distribution from the SM GF one.

To quantify the sensitivity of the LHC towards the xSM model, we perform a one dimensional binned log-likelihood ratio analysis, exploring the $m_{hh}$ distribution. In Fig.~\ref{figure:bound}, we display $2\sigma$ exclusion bound for the  $\mathcal{BR}(h_2\rightarrow h_1h_1)$ as a function of the heavy Higgs mass $m_{h_2}$. We assume the Higgs mixing $\sin\theta=0.2$. The high luminosity LHC with 3~ab$^{-1}$ will be sensitive  to ${\mathcal{BR}(h_2\rightarrow h_1h_1)<0.5}$ in the full mass range $300$~GeV$<m_{h_2}<1000$~GeV, using the $4b$ channel but a branching ratio as small as $\sim 0.05$ can be probed for heavy Higgs masses around 500 GeV.

We also overlay on this plot the region allowed by the theoretical 
potential requirements~\footnote{{
Some outliers exist above this region from
a more extensive scan of the parameter space, but is of negligibly
small parameter space, compared with the points falling within the
color regions. We also note that for this value of $\theta$, the region where $m_{h_2} \gtrsim 900\text{GeV}$ is right on the verge of being excluded by the $W$ boson mass constraint. However, we expect the parameter space of the detectable GW with a slightly smaller $\theta$, which evades the $W$ boson mass constraint, to have very
minor shift in this plot. So we choose to keep these points to show the feature of the parameter space.
}
}, obtained by a scan with $\sin\theta=0.2$
for each $m_{h_2}$.
Also shown here are points that can give a detectable GW signal at LISA, where green points
give $10< \text{SNR}< 50$ and red points gives $\text{SNR}> 50$. 
{To see the impact of the GW suppression on the SNR, we use $\delta=1$ in the left
panel and $\delta=0.01$ for the right panel. The reduction of GW production from sound
waves leads to a significant shrinking of the parameter space capable of generating detectable GW. However, the overall behavior of the remaining parameter space in affecting
the branching ratio remain qualitatively unaffected.
Compared with the theoretically allowed green region, this parameter space leads to an overall reduction of the di-Higgs branching ratio, especially for heavy $h_2$.
}
We can thus see clearly the complementary
role played by colliders and GW detectors in probing the xSM. For lighter $h_2$, i.e., 
$300~\text{GeV}\lesssim m_{h_2} \lesssim 600~\text{GeV}$, colliders will be able to probe, in the $4b$ channel alone, almost the entire parameter space that
is capable of generating a detectable GW signal. For heavier $h_2$, it is difficult for the HL-LHC to explore this regime because of phase space suppression. 
However, as the GW signal spectra with $m_{h_2}>600$~GeV present very small $\mathcal{BR}(h_2\rightarrow h_1h_1)$, a dedicated study of the $h_2\to WW,ZZ$ channel to determine the potential of the LHC to probe those points, beyond that performed in Ref.~\cite{Alves:2018oct}, might be interesting.

\section{\label{sec:summary}Summary}

Gravitational waves from the EWPT  provide a new window for probing the physics beyond the standard model,  complementing the current direct collider searches. We continue this complementarity study in this work by focusing on the di-Higgs production in the $4b$ channel, choosing the benchmark model xSM. In calculating the gravitational wave spectra, we carefully accounted for
several subtle issues, such as the bubble wall velocity, the supercooled phase transitions, and especially the  reduction in the gravitational wave production from sound waves outlined in recently conducted numerical simulations~\cite{Cutting:2019zws}.
These constitute important ingredients towards a faithful characterization of the GW signals from the EWPT and its detection at future gravitational wave detectors. The most important advance in recent understandings of these problems is the reduction of the GW produced from the sound waves, which 
invalidates the previous naive generalization of these formulae to arbitrary values of $v_w$ and $\alpha$. We incorporated this effect by applying
a conservative reduction factor of $0.01$.

In order to establish the complementary role of collider searches, we performed an analysis in the resolved $h_2\to h_1h_1\to 4b$ channel. We found that the 13 TeV HL-LHC is able to probe the xSM parameter space with $\mathcal{BR}(h_2\to h_1h_1)<0.5$  for $300~\text{GeV}<m_{h_2}< 1000$ GeV. 
It is clear from our analysis that due to the significant reduction of the gravitational wave signal strength, the xSM parameter space, which is capable of giving a detectable stochastic GW background, have shrunk. However, the qualitative complementarity role of future space-based GW detectors in assisting BSM physics searches at current and HL-LHC remains unchanged.

\begin{acknowledgments}
AA  thanks Conselho Nacional de Desenvolvimento Cient\'{i}fico (CNPq) for its financial support, grant 307265/2017-0.
DG was partially  supported by the U.S.~National Science Foundation under the grant PHY-1519175. 
TG is supported by U.S.~Department of Energy grant DE-SC0010504.
KS and HG are supported by the U.S. Department of Energy grant DE-SC0009956.
\end{acknowledgments}


\end{document}